\documentclass{ws-procs9x6}

\begin{document}

\title{A National Underground Science Laboratory in the U.S.}

\author{A.~B. BALANTEKIN\\
\lowercase{for the} NUSEL C\lowercase{ollaboration}
\footnote{\uppercase{F}or the full list of membership of the 
\uppercase{NUSEL C}ollaboration see the \uppercase{URL}  
``http://mocha.phys.washington.edu/\uppercase{NUSL}/collaboration.html$\#$membership''.}}

\address{University of Wisconsin, Department of Physics \\
Madison, WI  53706,  USA \\ 
E-mail: baha@nucth.physics.wisc.edu}

\maketitle

\abstracts{Recent developments in underground science are reviewed. 
Recent efforts to built a deep multi-purpose underground laboratory 
in the U.S. to explore a wide range of science are summarized.} 

\section{Underground Science}\label{1}

During recent years experiments done underground have produced some of 
the most significant new results in physics. In addition to ongoing research 
at the frontiers of nuclear physics, particle physics and astrophysics, 
underground science also includes some very exciting work going on 
in geology and biology, as 
well as potential to impact on applied areas such as materials science and 
nuclear proliferation. As we illustrate below the multidisciplinary 
research activity of underground experiments has the potential to make 
a major impact in the development of modern science. 

The field of underground physics has recently made fundamental discoveries 
that broadly impact physics, astronomy, and cosmology. Neutrinos coming from 
a core-collapse supernova were detected for the first time in 1987 with the 
observation of Supernova 1987A in the Large Magellanic Cloud. The same 
year, two-neutrino double beta decay was observed, which is the rarest process 
yet seen in Nature. In the 1990's variations of the atmospheric neutrino flux 
were observed for which the 
simplest explanation is neutrino oscillations due to the mixing of mass 
eigenstates. This result was the first demonstration of existence of physics 
beyond the standard model. The observation of the heavy-flavor neutrinos in 
the solar neutrino flux not 
only resolved the long-standing solar neutrino puzzle, but also provided an 
additional confirmation of the Standard Solar Model besides helioseismological 
observations. The pattern of neutrino masses emerging from these experiments 
implies new physics characterized by an energy scale $\sim 10^{15}$ GeV, many 
orders of magnitude beyond the direct reach of accelerators. The neutrino 
mixing angles emerging from the analysis of these experiments are close to 
maximal. This surprising result have consequences ranging from models of 
leptogenesis in the early universe to the explosion mechanism and 
nucleosynthesis in core-collapse supernovae. 

Another interesting recent development in underground science is the 
discovery of novel microorganisms that live deep in the earth. These 
microorganisms, which show an ability to adopt to extreme environments, 
provide an excellent opportunity for studying evolution in isolated 
environments. The young field of geomicrobiology explores how microbial 
life can exist in dark, hot, and high-pressure environments and the 
implications of its findings for the early development of life. These 
life-forms may be the most accessible model for astrobiology, searching 
for evolution, adaptation, and transport of life elsewhere. Deep underground 
mines also provide earth scientists and engineers prospects to study 
stability of large caverns, and thermal properties, hydrology, and ecology 
of deep rock. In applied science low-background environments of deep sites 
provide locations for the production and storage of materials free of 
cosmogenic activity, for the assessment of the effects of radiation damage 
on the performance of chips and other electronics, and for the establishment 
of multi-purpose low-background counting facilities. 

A recent committee of the National Research Council, Committee on the Physics 
of the Universe, listed eleven questions about the universe that identify 
most important and timely science opportunities at the intersection of physics 
and astronomy\cite{r0}. Underground science will be crucial in answering at
least six 
of these questions, namely i) What is dark matter? ii) How did the universe 
begin? iii) What are the masses of neutrinos, and how have they shaped the 
evolution of the universe? iv) Are protons stable? v) Are there new states 
of matter at exceedingly high density and temperature? vi) How are the 
elements from iron to uranium made? In general it is widely quoted that 
two great challenges of the science in the twenty-first century are exploring 
the nature of the dark matter/ dark energy and the origin and evolution of 
life. Underground science is likely to make a significant impact on both of 
these fields.

The next generation of underground experiments is more technically 
challenging than previous research requiring both significant resources and 
good planning and management to succeed. Establishment of a single 
deep multi-purpose 
underground laboratory will ensure that the needs of the diverse groups of 
physicists, earth scientists, geomicrobiologists, and others interested in 
underground science will be met. Such a laboratory will allow a common 
infrastructure (such as a facility to do low-background counting) 
to be be shared by various research groups and provide an environment where 
synergistic 
interactions among scientists working on diverse fields is possible. 
An underground science laboratory will also provide a critical mass of 
scientists and engineers for undertaking projects on education and outreach 
to the general public, both locally and nationally.

\section{Bahcall Committee}\label{2}

Current efforts to establish a deep-sited underground science laboratory in 
the U.S. started during the recent long-term planning exercise in nuclear 
science. In preparation for this planning exercise a group of neutrino 
physicists met in Seattle in September 2000 to discuss both the present state 
of the field and the new opportunities. At the same time it was announced that
the Homestake mine, where the Nobel-prize winning experiment of Davis and 
collaborators has been located, would be closed at the end of 2001. This 
provided an opportunity to create a deep underground multipurpose laboratory 
and impetus to move rather quickly to take advantage of this opportunity. The 
report summarizing conclusions of this meeting rated the creation of such a 
laboratory as the highest priority of this community\cite{r1}. 

Following 
the Seattle meeting a committee chaired by John Bahcall was formed to 
evaluate the scientific justification for a national facility for deep 
underground science and evaluate the suitability of suggested sites. 
This committee evaluated several U.S. sites including sites at Carlsbad (WIPP 
Facility), Homestake mine, and San Jacinto in Southern California. 
For comparison it also visited all major international facilities in Europe, 
Japan, and Canada. The depths of options considered are shown in the Table 1. 
The Bahcall Committee noted that that limited depth, and thus excessive 
cosmogenic background, preclude WIPP and Soudan sites from being general
purpose underground science facilities. (For 
a detailed discussion of the WIPP site see the contribution of 
Haines\cite{r2})
The committee endorsed a single 
primary site as the most effective method of realizing the anticipated 
scientific program and identified two deep sites, Homestake and San Jacinto, 
as two primary candidates. The depths of various existing underground 
laboratories as well as several options for the Homestake mine are 
shown in Figure 1. 

\begin{table}[t]
\tbl{Sites Considered by the Bahcall Committee}
{\footnotesize
\begin{tabular}{|c|r|}
\hline
{} &{}\\[-1.5ex]
Carlsbad (WIPP), NM  &   1600-2000 mwe\\[1ex]
Soudan,MN &              about 2000  mwe\\[1ex]
Homestake, SD &          up to 7200 mwe\\[1ex]
San Jacinto, CA &        5400-6300 mwe\\[1ex]
Greenfield sites (CA and NV) &  5000-8000 mwe\\[1ex]
\hline
\end{tabular}\label{tab1} }
\vspace*{-13pt}
\end{table}

\begin{figure}
\vspace{8pt}
%\figurebox{2cm}{3cm}{try}
%\includegraphics [width=3.5in,clip=yes]{depth2.eps}
\centerline{\epsfxsize=3.9in\epsfbox{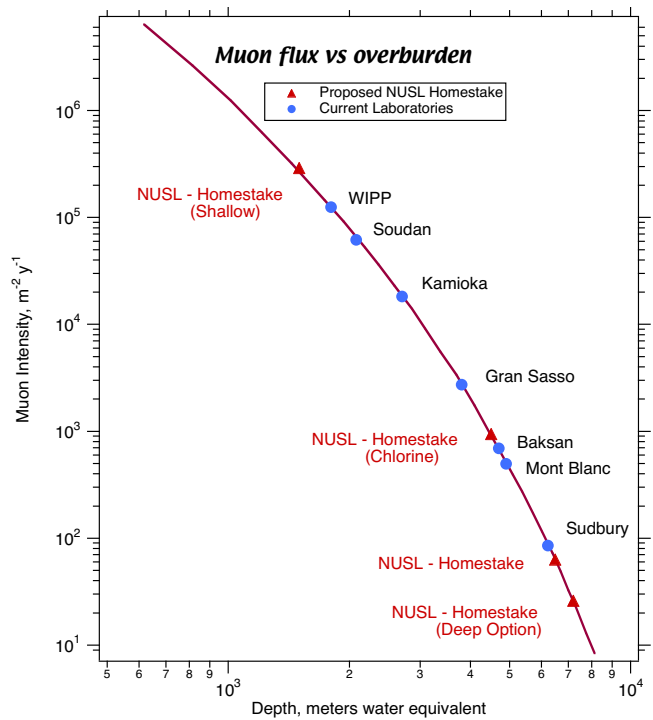}}   
\caption{Muon flux versus and depths of existing underground 
laboratories and various options for the Homestake mine.\label{fig1}}
\end{figure}

The Homestake site has the advantages of having i) Existing access to the 
proposed main level at 7400 ft; ii) Permissions in place 
for construction, waste rock 
disposal and safety; iii) Multiple access routes and venting for all levels; 
iv) Capacity to isolate risky experiments; and v) Well-characterized and 
understood rock mechanics. Although vertical access is 
often considered inconvenient, the Homestake facility has massive shafts and 
tunnels allowing efficient movement of very large loads to depth. 
The main issue with this 
site is the 
need to transfer the  
ownership of a privately-owned mine to the appropriate public venue. 
On the other hand the San Jacinto site has the advantages of i)  Construction 
permitting horizontal access; ii) Being in close proximity to major research 
centers in California; and iii) Likely lower operations cost. However it will 
take longer time to the placement of first experiments and permissions for 
mining, transporting, and disposing the waste rock 
have not been obtained. Another issue is the viability of the single tunnel 
design. In contrast to Homestake and tunnel facilities such as Gran Sasso, the
integrity of the San Jacinto site cannot be established prior to construction 
since the necessary core studies are not permitted because of federal 
ownership. 

\section{Towards a U.S. Underground Science Laboratory}\label{3}

It took another year after the Seattle meeting in 2000 for the long-range 
planning exercise for the nuclear physics to be completed. Citing a 
compelling opportunity for nuclear scientists to explore fundamental 
questions in neutrino physics and astrophysics, nuclear science community 
recommended immediate construction of a deep underground science 
laboratory\cite{r3}. Subsequently such a laboratory was strongly endorsed by 
the High Energy Physics Advisory Panel. The Committee on the Physics 
of the Universe, mentioned above, also recommended in its final report that an 
underground laboratory with sufficient infrastructure and depth be 
built\cite{r0}. The elaborations of the physics community 
was carried through a summer study in 2001 at Snowmass; two workshops (one 
on physics and one on geomicrobiology and Earth sciences) in the Fall of 2001 
at Lead, South Dakota; and an Aspen workshop in the summer of 2002.

Soon after the release of the report of the Bahcall committee teams 
representing both deep sites submitted 
proposals to the Federal funding agencies in the United States. Their web 
sites are {\tt http://mocha.phys.washington.edu/NUSL/} for Homestake and 
{\tt http://www.ps.uci.edu/~SJNUSL/} for San Jacinto. The team 
representing the Homestake site formed the NUSEL Collaboration. This is a 
group of scientists who work on a variety of tasks important to the creation 
of an underground laboratory at Homestake. 

The U.S. Federal budget request for the fiscal year 2003 asked  the National 
Science Council (NRC) to review the merits of several U.S. neutrino 
experiments in the context of the current and planned neutrino research 
capabilities through the world. NRC formed a committee to answer this charge.
The experiments mentioned in this charge roughly fall into two classes: Those 
experiments that study fundamental neutrino properties, dark matter, and 
relatively lower energy astrophysical sources (i.e. Sun, supernovae) require 
the very low-background environments of underground laboratories. Those 
experiments that study very high-energy neutrinos require large volumes of 
neutrino telescopes. The NRC committee is expected to address the unique 
capabilities of each class of new experiments and any possible redundancy 
between these two types of facilities that explore complementary, but 
scientifically very different questions. To help this committee and to 
produce a road map that will guide investigations worldwide over the
next few years a workshop took place in September 2002 at Washington, D.C. 
This workshop was designed to review and integrate the results of
recent developments in neutrino, low background and geoscience 
investigations requiring great subterranean depth. With the findings of this 
workshop and other results in hand the NRC committee is expected to release 
its final report by the end of 2002. 
%If the questions related to the 
%indemnification of the Homestake mine are resolved in a relatively timely 
%manner as well, 
The funding agency decisions are expected after the release 
of this committee's report.

%\section*{Acknowledgments}
This work was supported in part by the U.S. National Science
Foundation Grants No. PHY-0136261 and PHY-0070161. 
%, and in part by the 
%University of Wisconsin Research Committee with funds granted by the 
%Wisconsin Alumni Research Foundation.

\end{document}